\documentstyle[12pt,psfig]{article}

\newcommand{\norm}[1]{\raise.3ex\hbox{:} #1 \raise.3ex\hbox{:}\,}
\newcommand{\disc}{{\rm Disc}}

\def\TL{\hfil$\displaystyle{##}$}
\def\TR{$\displaystyle{{}##}$\hfil}

\def\comment#1{}
\def\fixit#1{}

\def\tf#1#2{{\textstyle{#1 \over #2}}}

\def\tr{\mathop{\rm tr}\nolimits}

\def\lsim{\mathrel{\mathstrut\smash{\ooalign{\raise2.5pt\hbox{$<$}\cr\lower2.5pt\hbox{$\sim$}}}}}
\def\gsim{\mathrel{\mathstrut\smash{\ooalign{\raise2.5pt\hbox{$>$}\cr\lower2.5pt\hbox{$\sim$}}}}}

\def\sqr#1#2{{\vcenter{\vbox{\hrule height.#2pt
         \hbox{\vrule width.#2pt height#1pt \kern#1pt
            \vrule width.#2pt}
         \hrule height.#2pt}}}}

\def\href#1#2{#2}  

\def\eqn#1#2{\begin{equation} #2 \label{#1} \end{equation}}
\def\eqalign#1{\vcenter{\openup1\jot
    \halign{\strut\span\TL & \span\TR\cr #1 \cr
   }}}

\begin{document}
\begin{titlepage}

\begin{flushright}
PUPT-1770\\
NSF-ITP-98-022\\
hep-th/9803023
\end{flushright}
\vfil\vfil

\begin{center}

{\Large {\bf Scalar Absorption and the Breaking of 
the World Volume Conformal Invariance}}

\vfil

Steven S.~Gubser$^{a,b}$,  Akikazu Hashimoto$^b$, 
Igor R.~Klebanov$^{a,b}$, and  Michael Krasnitz$^a$

\vfil

$^a$Joseph Henry Laboratories\\
 Princeton University\\
 Princeton, New Jersey 08544\\

\vfil

$^b$Institute for Theoretical Physics\\
 University of California\\
 Santa Barbara, CA  93106
\end{center}


\vspace{5mm}

\begin{abstract}
\noindent 
We investigate a version of fixed scalars for non-dilatonic branes
which correspond to dilatations of the brane world-volume.  We obtain
a cross-section whose world-volume interpretation falls out naturally
from an investigation of the breaking of conformal invariance by the
irrelevant Born-Infeld corrections to Yang-Mills theory.  From the
same irrelevant world-volume operator we obtain the leading correction
to the cross-sections of minimal scalars.  This correction can be
obtained in supergravity via an improved matching of inner and outer
solutions to the minimal wave equation.
\end{abstract}

\vfil\vfil\vfil
\begin{flushleft}
March 1998
\end{flushleft}
\end{titlepage}

\newpage
\renewcommand{\baselinestretch}{1.05}  

\section{Introduction}
\label{Intro}

Recently there has been major progress in understanding the strong
coupling limit of large $N$ gauge theory
\cite{gkp,kleb,gukt,gkThree,Sasha,jthroat,US,ho,EW,ks,Ferrara,Juan,vafa}.
Maldacena's work \cite{jthroat} has connected the applications of the
large $N$ limit of coincident D3 branes to black 3-branes
\cite{gkp,kleb,gukt,gkThree} with Polyakov's sigma model approach to
confining strings in large $N$ gauge theories \cite{Sasha}.  The
conjecture of \cite{jthroat} relates the world volume theories of
certain coincident branes to superstring or M-theory on the
backgrounds consisting of anti-deSitter ($AdS$) spaces times spheres,
which arise as the throat limits of the relevant black brane
geometries \cite{gt,ght}.

Earlier calculations revealing connections between correlators in the
world volume theories and low-energy absorption in the entire brane
metric may be found in \cite{kleb,gukt,gkThree}. Taking the throat
limit identifies the ``universal'' part of the metric which determines
the correlation functions in the conformal limit \cite{jthroat,US,EW}.
Indeed, because the isometry groups of $AdS$ spaces are conformal
groups, it is sensible for them to be associated with a conformal
theory on the world-volume.  However, brane geometries are only $AdS$
out to a characteristic radius $R$ which scales as a power of the
number of branes.  Furthermore, when one specifies the world-volume
theory with the DBI action, powers of the string scale $\alpha'$
suppress the non-renormalizable interactions.  We find that, even for
large $N$ and strong coupling, when the entire brane geometry is
smooth \cite{kleb}, these corrections are easily detectable in the
energy dependence of the absorption cross-sections. The purpose of
this paper is to identify and explain these effects.

To examine the departures from conformal invariance more closely, we
propose two concrete probes. On the one hand, we examine the
calculations of absorption of minimally coupled scalars \cite{kleb} in
more detail.  The wave equation for minimal scalars propagating in the
extremal 3-brane background is given by
\eqn{minmassless}{\left
 [\rho^{-5} {d\over d\rho} \rho^5 {d\over d\rho} +
1 + {(\omega R)^4\over \rho^4} \right ] \phi(\rho) =0
\ .}
The coupling to the background geometry is controlled by dimensionless
parameter 
$$(\omega R)^4\sim \omega^4 \kappa N \ ,$$  
where the gravitational constant $\kappa \sim g_{YM}^2 (\alpha')^2$.
In \cite{kleb} a general relation was proposed between a certain limit
of the type IIB supergravity in the self-dual 3-brane background, and
the corresponding limit of the world volume gauge theory. This is the
double-scaling limit,
\eqn{dsl}{
g_{YM}^2 N \rightarrow \infty\ , \qquad\qquad
\omega^2\alpha' \rightarrow 0\ ,
}
where $\omega^4 (\alpha')^2 g_{YM}^2 N\sim (\omega R)^4$ is kept fixed
and small.  One expects the absorption cross-section to contain
corrections of higher order in this parameter:\footnote{A similar
point was discussed in \cite{Das:1997kt}.}
$$ \sigma  \sim \kappa^2 N^2 \omega^3 
\left(1 + c_1 (\omega R)^4 + c_2 (\omega R)^8 + \ldots \right)$$
In this paper
we present a systematic procedure for obtaining such corrections to
the absorption cross-section.  Somewhat surprisingly,
we find that the leading correction contains  a logarithmic term
$$\sigma  \sim \kappa^2 N^2 \omega^3 \left(
1 + b_1' (\omega R)^4 \log (\omega R) + b_1 (\omega R)^4 + \ldots
\right )\ ,$$
which dominates in the small $\omega R$ limit over the $(\omega R)^4$
term. Nevertheless, this expansion is consistent with the 
double-scaling limit (\ref{dsl}).
The logarithmic term encodes the leading departure from the
conformal limit.  For the
D3-branes, this breakdown is naturally interpreted in terms of the DBI
action: at lowest order, the DBI action reduces to ${\cal N}=4$ SYM
theory, which is conformal in $4$ dimensions; but as one proceeds to
higher orders, non-renormalizable interactions enter into the
Lagrangian. We argue
that the $\tr F^4$ term from the DBI action \cite{Arkold,gw,Ark}
in fact reproduces the
peculiar logarithmic form of the leading correction to the absorption
probability.

The other probe is a higher-dimensional analogue of the fixed scalars
considered in four and five dimensions
\cite{kr,cgkt,kkOne,kkTwo,Lee:1997xz,Lee:1997xg}.  In a compactified
geometry where the brane is wrapped around a torus, the scalar we
consider is the volume of the torus.  Variations of this scalar
correspond to dilatations of the world volume; hence it couples to the
trace of the stress-energy tensor.  This trace vanishes for a
conformal theory, and correspondingly the absorption of the fixed
scalar is suppressed by $(\omega R)^4$ where $\omega$ is the energy.
It turns out for the D3-brane that the same $\tr F^4$ term in the DBI
action gives a contribution to the stress-energy tensor which explains
the absorption probability at leading order in low energies.

\section{Fixed Scalars and Conformal Operators}

It is by now well-known that, in the background of certain
supersymmetric black holes in 4 and 5 dimensions, there exist
non-minimally coupled massless scalar fields. One class of such
fields, called the fixed scalars, was discovered in
\cite{fk,fkone,fks}.\footnote{ For $D=5$ there exists yet another
class, called the intermediate scalars \cite{KRT}.}  Due to the
non-minimal couplings, the low-energy absorption cross-sections for
such fields are suppressed compared to those of the minimally coupled
scalars \cite{kr,cgkt,kkOne,kkTwo}. In the effective string models of
$D=4$ and $D=5$ black holes this suppression has a natural
explanation: while the minimal scalars couple to marginal operators,
the non-minimal ones couple to irrelevant operators
\cite{cgkt,kkOne,kkTwo}. Such operators are ignored in the conformal
limit, but are well-known to be present in the non-polynomial actions
of the DBI type.

In this section we calculate the absorption cross-sections of fixed
scalars by charged black holes in $D>5$, and find the suppression at
low energies similar to that discovered in $D=4$ and $5$. Particularly
interesting from our point of view are the cases $D=7$, $6$ and $9$,
which are related to multiply wrapped D3, M5 and M2 branes
respectively. These are precisely the theories that have been
receiving much attention recently, and the fixed scalars provide a
novel way of probing them.

The semiclassical absorption cross-sections of minimally coupled
scalars (such as the dilatons) were studied in
\cite{kleb,gukt,gkThree} and were found to be consistent with world
volume considerations.  For instance, the leading coupling of the
dilaton to the $U(N)$ theory on $N$ coincident D3-branes is
\begin{equation}
\int d^4 x \, \phi (x) {\cal O}_4
\end{equation}
where ${\cal O}_4$ is the marginal operator which changes $g_{YM}$,
\eqn{dimfour}{
{\cal O}_4=\tr F^2+ \ldots \ . }
At low energies the absorption cross-section is found to behave as $
\sigma \sim \kappa^2 N^2 \omega^3$, consistent with the fact that the
exact dimension of ${\cal O}_4$ is equal to 4.  Other fields that act
as scalars from the point of view of the $D=7$ black hole include the
gravitons polarized in the internal dimensions (along the D3
branes). Their world volume coupling is given by
\begin{equation}
\int d^4 x \, {1\over 2} h_{\mu\nu}  T^{\mu\nu}
\end{equation}
where $T^{\mu\nu}$ is the stress-energy tensor.  Now, we may consider
a particular scalar field, $h^\mu_\mu$.  Since its vertex operator is
$T^\mu_\mu$, this field obviously decouples in the conformal
limit. This is precisely the property expected of a fixed scalar, and
we will show that the corresponding linearized equation indeed
contains a non-minimal term.

We have concluded that, for D3 branes, $h^\mu_\mu$ does not couple to
a marginal (dimension 4) operator. We may deduce the leading operator
which it couples to from the well-known structure of the DBI action
\cite{Arkold,gw,Ark}. To order $F^4$, we have
\begin{equation}
S_{\rm DBI}= {1\over 4 g_{YM}^2 }\int d^4 x \, \left [
 \tr F_{\mu\nu}^2 - 2 (\pi \alpha')^2 {\cal O}_8 + 
\ldots \right ]
\ .\end{equation}
The operator\footnote{We have not exhibited the dependence of this
operator on the scalars and the fermions. We believe that these extra
terms are determined by supersymmetry.}
\eqn{irrop}{
{\cal O}_8= {2\over 3} \tr 
\left (F_{\mu\nu} F_{\rho\nu}
F_{\mu\lambda} F_{\rho\lambda}+
{1\over 2} F_{\mu\nu} F_{\rho\nu} F_{\rho\lambda}F_{\mu\lambda}
-{1\over 4} F_{\mu\nu} F_{\mu\nu} F_{\rho\lambda} F_{\rho\lambda}
-{1\over 8} F_{\mu\nu} F_{\rho\lambda} F_{\mu\nu} F_{\rho\lambda}
\right )}
has bare dimension 8 and obviously breaks conformal invariance. Thus,
the trace of the stress-energy tensor calculated from this term is
also of dimension 8, i.e. the lowest dimension coupling of the fixed
scalar to the world volume is of the form
\begin{equation}
\int d^4 x \, h^\mu_\mu {\cal O}_8 (2\pi \alpha')^2 
\ .\end{equation}
The leading contribution to the 2-point function is a 3-loop diagram,
which scales as
\begin{equation}
\langle {\cal O}_8 (x) {\cal O}_8 (0) \rangle \sim {N^2 \left (N g_{YM}^2
\right )^2
\over x^{16}}
\ .
\end{equation}
In fact, our absorption calculations will give us reasons to believe
that this formula is true non-perturbatively, i.e. this operator does
not receive any anomalous dimension in the CFT. Assuming this,
performing the Fourier transform and isolating the imaginary part, we
find that the absorption cross-section should behave as
\begin{equation}
\sigma \sim N^4 \kappa^4 \omega^{11}
\ .\end{equation}
In the next subsection we will see that precisely this scaling results
from the semiclassical gravity calculation.

A parallel analysis may be performed also for the coincident M5 and M2
branes. In both cases we find that trace of the 11 dimensional
graviton over indices parallel to the brane is the fixed scalar field,
and its leading order coupling is to a relevant operator of twice the
marginal dimension. Of course, in these cases comparison with the
world volume considerations is weaker since the theory of multiple
coincident branes is poorly understood.

\subsection{Semiclassical absorption of fixed scalars}

We obtain a charged black hole in $D=10-p$ by wrapping some number of
Dirichlet $p$-branes over $T^p$.  The part of the $D$-dimensional
effective action that will be relevant for our calculations is
\cite{hs} \eqn{action}{ \int d^D x \sqrt{-g}\left ( R - {1\over 2}
\partial_m \lambda\partial^m \lambda- e^{\beta \lambda} G_{mn} G^{mn}
\right ) \ , } where
$$\beta = \sqrt {2 (D-1)\over D-2}
\ ,
$$
and $G_{mn}$ is the $U(1)$ field strength.  The fixed scalar $\lambda$
is a certain linear combination of $\log V$ and the 10-dimensional
dilaton $\phi$ ($V$ is the internal volume of the brane measured in
the 10-dimensional Einstein metric), 
\eqn{mix}{ \beta \lambda = {D-7 \over 2} \phi - \left( {2 D-6 \over
D-2} \right) \log V \ .  }

The static charged black hole solution is
\eqn{sol} {
d s^2 = A^{1\over D-2} (- A^{-1} dt^2 + dr^2 + r^2 d\Omega_{D-2}^2 )
\ ,
}
$$ G_{rt} = {1\over\sqrt 2} \partial_r A^{-1}\ ,
\qquad\qquad \lambda= {\beta\over 2} \log A
\ ,$$
$$A(r) = 1+ {R^{D-3}\over r^{D-3} }= 1 + 
{2Q \over (D-3)r^{D-3}}\ .
$$
We will see that perturbing about this solution works similarly to the
$D=5$ case which has already been worked out for the triply charged
black hole. In fact, setting two of the three charges to zero in
\cite{kkTwo} we may immediately obtain the fixed scalar fluctuation
equation for $D=5$. This is one of the consistency checks on our new
calculations. In fact, we will be able to work out the equations for a
general $D$.  The cases of special interest to us are D=7,
corresponding to coincident D3 branes in 10 dimensions; D=6,
corresponding to coincident M5 branes in 11 dimensions or D4 brane in
10 dimensions; and D=9, corresponding to coincident M2 branes in 11
dimensions or strings in 10 dimensions.

One may be concerned that the $D=7$ case should be treated separately
because the solution also includes the 5-form background $H= \star G$.
Thus, a priori the action is
\eqn{otheract}{\int d^7 x \, \sqrt{-g}\left ( R - {1\over 2} \partial_m
\lambda\partial^m \lambda - e^{\beta \lambda} G_{mn}
G^{mn}- {2\over 5!} e^{-\beta \lambda}
H_{m_1 \ldots m_5} H^{m_1 \ldots m_5} \right )
\ .
}
However, we may dualize the $H^2$ term into the $G^2$ term, so that
the action is equivalent to
$$ \int d^7 x \, \sqrt{-g}\left ( R - {1\over 2} \partial_m
\lambda \partial^m \lambda -2 e^{\beta \lambda} G_{mn}
G^{mn} \right )
\ .
$$
This makes it look essentially the same as the problem in $D\neq 7$.
The extra factor of 2 in front of the $G^2$ term is compensated by the
fact that the classical electric field has an extra $1/\sqrt 2$: in
this case
$$ G_{rt} ={1\over 2} \partial_r A^{-1}\ .
$$

In studying the propagation of fixed scalars, special care needs to be
taken to account for mixing with the gravitational field.  This mixing
can be traced to the fact that $\lambda$ couples to background
electric field of the black hole. Fortunately, the methods for
disentangling this mixing have been developed in \cite{kkTwo}. Their
application here is a straightforward generalization of these methods,
and we just summarize the results.

The general formula for the fixed scalar potential is
\eqn{genpot}{
{2(D-1)^2(D-3)^2Q^2 \over r^2[(D-1)Q+(D-2)(D-3)r^{D-3}]^2} \ .
}
Using
$R^{D-3} = {2Q \over D-3}$, we find that
the fixed scalar fluctuations in D dimensions obey the equation
\eqn{genfluct}{
\left [r^{-(D-2)}\partial_rr^{D-2}\partial_r+\omega^2 \left (1+{R^{D-3}
\over r^{D-3}}\right )-{2(D-1)^2(D-3)^2R^{2(D-3)}
\over r^2[(D-1)R^{D-3}+2(D-2)r^{D-3}]^2}
\right ]\lambda = 0\ .}

It remains to find an approximate solution of this equation for low
energies and derive the absorption cross-section.  As in previous work
\cite{kr,cgkt,kkOne,kkTwo}, we divide space into three regions and
match.  In the near region the equation is
\eqn{near}{
\left [\rho ^{-(D-2)}d_{\rho}\rho ^{D-2}d_{\rho}+
{(\omega R)^{D-3}\over \rho^{D-3}}-{2(D-3)^2 \over \rho^2}
\right ]\lambda_I=0\ .}
where $\rho = \omega r$. Letting 
\eqn{letting}{\rho =  Az^{2/(5-D)}\ ,\qquad\qquad 
A^{D-5} = {4(\omega R)^{D-3}\over (D-5)^2}\ ,}
we find
$$\lambda_I = z^{(D-3)/(D-5)}H_{\nu}(z)$$
where
$$\nu = {3(D-3)\over D-5}\ .$$

In the intermediate region the $\omega$ term is irrelevant and we get
the solution
\eqn{interm}{
\lambda_{II}(r) = {Br^{D-3}\over (D-1)R^{D-3} +2(D-2)r^{D-3}}\ .
}
Matching with the near region, we obtain
$$B = [{4 \over (D-5)^2}]^{(3-D)/(D-5)}
{(D-1)\Gamma(\nu)2^{\nu}\over \pi}(\omega R)^{-2(D-3)/(D-5)}\ .$$

In the far region we have the equation
$$\left [r^{-(D-2)}\partial_rr^{D-2}
\partial_r+\omega^2 \right ]\lambda_{III}=0 \ .$$
Its solution is
$$\lambda_{III} = C\rho^{-\mu}J_{\mu}(\rho),$$
where $\mu = (D-3)/2$. Matching to the intermediate region, we get
$$C = {2^{\mu}\Gamma(\mu+1)\over 2(D-2)}B \ .$$

The invariant flux is given by
$(1/2i)(\lambda^*\partial_rr^{D-2}\lambda - c.c.)$.  Taking the ratio
of the flux at the horizon to the the incoming part of the flux at
infinity, we get the absorption probability
\eqn{absprob}{P = {4 \over |C|^2}{(D-5)\over 2}A^{D-3}\ ,}
which translates into
\eqn{absprobnew}{P = (D-5)
\left [{4\over (D-5)^2} \right ]^{3(D-3)/(D-5)}
{(D-2)^2\over (D-1)^2} { 8\pi^2 \over 2^{2(\mu+\nu)}
(\Gamma(\nu))^2(\Gamma(\mu+1))^2}(\omega R)^{(D-3)(D+1)\over D-5}\ .
}
The s-wave absorption cross-section is given by
$$\sigma= {(2\sqrt \pi)^{D-3} 
\Gamma\left({D-1\over 2}\right ) \over \omega^{D-2}} P
\ .
$$
Thus, we find
\eqn{abscross}{\sigma=  
(D-5)[{4\over (D-5)^2}]^{3(D-3)/(D-5)}
{(D-2)^2\over (D-1)^2} { 8\pi^2 (2\sqrt \pi)^{D-3}\over 2^{2(\mu+\nu)}
(\Gamma(\nu))^2 \Gamma(\mu+1)} R^{(D-3)(D+1)\over D-5}
\omega^{5D-13\over D-5}\ .
}
Let us exhibit the scaling of the cross-section with the number of
branes and the energy:
$$\sigma \sim N^{D+1\over D-5} \omega^{5D-13\over D-5}
\ .
$$

For $N$ coincident D3 branes, which correspond to the $D=7$ black
hole, we find
$$ \sigma_{D3} \sim \kappa_{10}^4 N^4 \omega^{11} \ .$$ 
As explained in the preceding section, this scaling shows that the
exact dimension of the operator ${\cal O}_8$, which the fixed scalar
couples to, is equal to 8. Thus, its anomalous dimension vanishes.  We
have, therefore, found another situation where gravity gives us a
``proof'' of a non-renormalization theorem for an operator in the
world volume theory.  We believe that in the gauge theory this theorem
follows from the existence of the supersymmetric DBI action, and from
the fact that insertions of ${\cal O}_8$ can be obtained by
differentiating the path integral with respect to $\alpha'$.

While our route towards the operator ${\cal O}_8$ involved using the
DBI action, which breaks the conformal invariance, the operator itself
is expected to be one of the chiral operators of the ${\cal N}=4$ SYM
theory.\footnote{We thank O. Aharony, H. Ooguri and J. Maldacena for
emphasizing this to us.}  This is required by the statement that the
chiral operators are in one to one correspondence with the massless
modes of type IIB supergravity \cite{US,ho,EW}.  Chiral operators
involving $\tr F^4$ have indeed been found in \cite{ds}. From the form
of the fixed scalar equation in the throat region, (\ref{near}), we
find that the AdS mass-squared of the corresponding state is
$$ m^2 = 32/ R^2\ .
$$
Thus, we believe that ${\cal O}_8$ should be identified with the $k=0$
(the $SO(6)$ singlet) state in the tower
\eqn{tower} {
m^2 = (k+4) (k+8)/R^2\ ,
}
which appears in type IIB supergravity on $AdS_5\times S^5$ \cite{Kim}.

For $N$ coincident M5 branes, which correspond to the $D=6$ black
hole, we find
$$ \sigma_{M5} \sim \kappa_{11}^{14/3} N^7 \omega^{17}
\ .$$
This indicates that the fixed scalar couples to an operator of
dimension 12 on the 6-dimensional world volume.

For $N$ coincident M2 branes, which correspond to the $D=9$ black
hole, we find
$$ \sigma_{M2} \sim \kappa_{11}^{10/3} N^{5/2} \omega^8 \ .$$
This indicates that the fixed scalar couples to an operator of
dimension 6 on the 3-dimensional world volume.  It is also known that
in the effective string models describing black holes in $D=4$ and
$5$, the fixed scalars couple to dimension 4 operators
\cite{cgkt,kkOne,kkTwo}. Thus, in all known cases we find that the
leading relevant term in the non-linear action, which couples to the
fixed scalar, has twice the marginal dimension.

\section{Higher order corrections to minimal scalar coupling}
\label{ImproveMatch}

In this section we will exhibit a technique for obtaining systematic
corrections to the minimal scalar cross-section.  We will start with
the D3-brane in section~\ref{ThreeImp}, then proceed to the M5-brane
in section~\ref{FiveImp}.  We will find it useful that the minimal
scalar equation in the 3-brane background has an inversion symmetry
which interchanges the throat and the asymptotic regions. There is
also a similar symmetry which interchanges the throat of the M2-brane
with the asymptotic region of the M5-brane and vice versa.  This
guarantees, among other things, that the absorption probabilities for
the M2 and M5 cases are identical.

Let us start with exhibiting this inversion symmetry for the extremal
3-brane.  The minimal massless scalar equation is given in
(\ref{minmassless}).  If we make a substitution, $\phi(\rho)=
\rho^{-2} \chi(\rho)$, then
$$\left  [ {d^2\over d\rho^2} + {1\over \rho} {d\over d\rho} +
1- {4\over \rho^2} + {(\omega R)^4\over \rho^4} \right ] \chi =0
\ .$$
On the other hand, if we introduce a variable $y=(\omega R)^2/\rho$,
then
$$ \left [ {d^2\over d y ^2} + {1\over y} {d\over d y} +
1- {4\over y^2} + {(\omega R)^4\over y^4} \right ] \chi =0
\ ,$$
which is the same equation! So, there is a inversion symmetry that
interchanges the $AdS$ and the asymptotic regions.

This symmetry is even evident in the metric.  If we start with the
extremal metric
$$ ds^2 = A(r)^{-1/2} dx_{||}^2 + A(r)^{1/2} (dr^2 + r^2 d\Omega_5^2)
\ ,$$
$$
A(r)= 1+{R^4\over r^4}
\ ,
$$
and define a new coordinate, $z= R^2/r$, then the metric becomes
$$ ds^2= {R^2\over z^2} \left [A(z)^{-1/2} dx_{||}^2 + 
A(z)^{1/2} (dz^2 + z^2 d\Omega_5^2) \right ]
\ .$$
So, after a conformal transformation the metric looks the same in
terms of $z$.

Now consider the massless equation in the background of a M5-brane,
\eqn{minfive}{\left [
\rho^{-4} {d\over d\rho} \rho^4 {d\over d\rho} +
1 + {(\omega R)^3\over \rho^3} \right ] \phi(\rho) =0
\ .}
If we make a substitution, $\phi(\rho)= \rho^{-3/2} \psi(\rho)$, then
$$\left [ {d^2\over d\rho^2} + {1\over \rho} {d\over d\rho} +
1- {9\over 4 \rho^2} + {(\omega R)^3\over \rho^3}
\right ] \psi(\rho) =0
\ .$$
Here the absorption probability is known to be \cite{kleb}
$$P = {\pi\over 9} (\omega R)^9\ .$$

For the M2-brane the equation is
\eqn{mintwo}{\left [\rho^{-7} {d\over d\rho} \rho^7 {d\over d\rho} +
1 + {(\omega R)^6\over \rho^6} \right ] \phi(\rho) =0
\ .}
Using the variable $y=(\omega R)^3/(2\rho^2)$, and setting $\phi =
y^{3/2} \chi$, one gets
$$\left [ {d^2\over d y^2} + {1\over y} {d\over dy} +
1- {9\over 4 y^2} + {(\omega R)^3\over 8 y^3} \right ] \chi(y) =0
\ ,$$
which is the same equation as for the M5-brane, except $\omega R
\rightarrow \omega R/2$. So, the outer region of the M2-brane is
mapped into the inner region of the M5-brane and vice versa.  The
tunneling probability between the two regions must be the same
function as for the M5 brane. So, here we find
$$P = {\pi\over 9} (\omega R/2)^9\ ,$$
which agrees with the direct calculation \cite{emp}.

Now, let us look at the metric. The M2-brane metric is
$$ ds^2 = A(r)^{-2/3} dx_{||}^2 + A(r)^{1/3} (dr^2 + r^2 d\Omega_7^2)
\ ,$$
$$
A(r)= 1+{R^6\over r^6}
\ ,
$$
Defining the coordinate $z=R^3/(2 r^2)$, we find
$$ ds^2 = {R^2\over 4 z^2}
\left [B(z)^{-2/3} dx_{||}^2 + B(z)^{1/3} (dz^2 + 4 z^2 d\Omega_7^2)
\right ]
\ ,$$
where
$$
B(z)= 1+{R^3\over 8 z^3}
\ .$$
After a conformal transformation this becomes
$$ ds_{new}^2 = 
B(z)^{-1/3} dx_{||}^2 + B(z)^{2/3} (dz^2 + 4 z^2 d\Omega_7^2)
\ .$$
The $rt$ part of this metric coincides with that of the M5-brane.

Now we proceed to calculating corrections to the low energy absorption
cross-sections.

\subsection{The D3-brane case}
\label{ThreeImp}

The equation of motion for a minimal scalar $\phi$ in the $s$-wave
is (\ref{minmassless}). In terms of 
$y = (\omega R)^2/\rho$ and $\phi = y^4 \psi$, we have
  \eqn{InnerDiffEQ}{
   \left[ {1 \over y^5} \partial_y y^5 \partial_y + 
    1 + {(\omega R)^4 \over y^4} \right] \psi = 0 
  \ .}
The strategy we will employ is to expand 
  \eqn{ExpandOuter}{
   \phi(\rho) = \phi_0(\rho) + (\omega R)^4 \phi_1(\rho) + \ldots
  }
 in the outer region where the last term in (\ref{minmassless}) is
small.  Writing down (\ref{minmassless}) order by order in $(\omega
R)^4$, we find
  \eqn{FirstTwo}{\eqalign{
   \left[ {1 \over \rho^5} \partial_\rho \rho^5 \partial_\rho + 
    1 \right] \phi_0 &= 0  \cr
   \left[ {1 \over \rho^5} \partial_\rho \rho^5 \partial_\rho + 
    1 \right] \phi_1 &= -{1 \over \rho^4} \phi_0 \ .  
  }}
 The two solutions to the homogeneous equation for $\phi_0$ are 
  \eqn{TwoSolns}{\eqalign{
   {J_2(\rho) \over \rho^2} &= 
    \tf{1}{8} \left( 1 - {\rho^2 \over 12} + O(\rho^4) \right)  \cr
   {N_2(\rho) \over \rho^2} &=
    -{4 \over \pi \rho^4} \left( 1 + {\rho^2 \over 4} + 
     O(\rho^4 \log\rho) \right) \ .  
  }}
 Using a trick of second order differential equations, we can write
down a solution of the inhomogeneous equation directly:
  \eqn{PartSoln}{
   \phi_1(\rho) = -{\pi \over 2 \rho^2} \int^\rho
    {d\sigma \over \sigma} \phi_0(\sigma) 
     \left( J_2(\sigma) N_2(\rho) - N_2(\sigma) J_2(\rho) \right) \ .
  }
 A convenient notation is to refer to $\phi_1$ as $\Delta \phi_0$.
Note however that $\phi_1$ is ambiguous because one can add to it any
solution to the homogeneous equation.  This ambiguity is fixed by
imposing the boundary conditions that 1) all flux at the horizon is
infalling, and 2) the solution of inner and outer region match to
order $(\omega R)^4$ in the transition region.\footnote{As we will see below,
the analysis of matching at the transition region will reveal a
dominant correction of order $(\omega R)^4 \log(\omega R)$ in the flux
ratio, and if we are only interested in computing the absorption cross
section to this order and not to order $(\omega R)^4$, we will be
justified in ignoring the effect of homogeneous terms of order
$(\omega R)^4$ on the asymptotic flux.  
}

The corrected solutions for the inner region ${\bf I}$ and outer
region ${\bf III}$ are as follows:
  \eqn{CorSolns}{\eqalign{
   \phi^{\bf I} &= y^4 (\psi_0 + (\omega R)^4 \psi_1)  \cr
     &= y^2 H_2^{(1)}(y) + (\omega R)^4 {\pi y^2 \over 2} 
       \int^y {dx \over x^3} H_2^{(1)}(x) 
       \left( J_2(x) N_2(y) - N_2(x) J_2(y) \right)  \cr
   {\phi^{\bf III} \over A} &= \phi_0 + (\omega R)^4 \phi_1  \cr
     &= {J_2(\rho) \over \rho^2} - (\omega R)^4 {\pi \over 2 \rho^2} \left[
       \int^\rho {d\sigma \over \sigma^3} J_2(\sigma)^2 N_2(\rho)  
 -        \int^\rho  {d\sigma \over \sigma^3} 
        J_2(\sigma) N_2(\sigma) J_2(\rho) \right]\ .  
  }}
By adjusting the contribution from the homogeneous terms of order
$(\omega R)^4$,  the wave function in the overlapping region 
can be made to match  up to order $(\omega R)^4 \log(\omega R)$
  \eqn{ExpSolns}{\eqalign{
   \phi^{\bf I} &= -{4 i \over \pi} \left[
    1 + {y^2 \over 4} - {(\omega R)^4 \over 12 y^2}
     \left( 1 - {y^2 \log y \over 2} \right) \right] +
      \ldots  \cr
    &= -{4 i \over \pi} \left[ 1 + 
      (\omega R)^2 \left( {1 \over 4 z^2} - {z^2 \over 12} \right) + 
      (\omega R)^4 \left( -\tf{1}{24} \log z + 
       \tf{1}{24} \log \omega R \right) + 
      \ldots \right]  \cr
   {\phi^{\bf III} \over A} &= \tf{1}{8} 
     \left( 1 - {\rho^2 \over 12} \right) + 
     {(\omega R)^4 \over 32 \rho^2} \left( 1 - 
      \tf{1}{6} \rho^2 \log \rho  \right) + \ldots  \cr
    &= \tf{1}{8} \left[ 1 + 
      (\omega R)^2 \left( {1 \over 4 z^2} - {z^2 \over 12} \right) + 
      (\omega R)^4 \left( -\tf{1}{24} \log z -
       \tf{1}{24} \log \omega R  \right) + 
      \ldots \right]  
  }}
 where we have introduced a new radial variable $z$:
  \eqn{ZDef}{
   z = r/R \qquad\ \rho = (\omega R) z \qquad\ y = {\omega R \over z} \ .
  }
The self-dual point is $z=1$.  The idea is to keep $z$ finite in the
matching region, so that both $\rho$ and $y$ are small.  Indeed,  we observe a leading mismatch at order $(\omega R)^4 \log(\omega R)$. If in
(\ref{ExpSolns}) we now choose 
  \eqn{AChoice}{
   A = -{32 i \over \pi} \left[ 1 +
    {(\omega R)^4 \over 12} \log \omega R - O((\omega R)^4)\right]
  }
then $\phi^{\bf I}$ and $\phi^{\bf III}$ match perfectly.  The
cross-section goes as $1/|A|^2$, so the corrected result is
  \eqn{CrossSRatio}{
   \sigma = {\pi^4 \over 8} \omega^3 R^8 
     \left( 1 - {(\omega R)^4 \over 6} \log \omega R + O((\omega R)^4)
     \right) \ .
  }
 We have verified the coefficient on the logarithm by numerically
solving the radial equation (\ref{minmassless}).
Note however that one has to get to very small $\omega R$ to see 
the logarithm numerically.

\subsection{The M5-brane case}
\label{FiveImp}

The methods for this case are the same as for the D3-brane, except in
that it is necessary to iterate the outer region perturbation twice.
The radial equation for a minimal scalar in the $s$-wave is
(\ref{minfive}). Defining 
  \eqn{VarDefs}{
   y = {2 (\omega R)^{3/2} \over \sqrt{\rho}} \qquad
   \phi = y^3 \psi \ .
  }
we also have
\eqn{MfiveEQ}{\left[ {1 \over y^7} \partial_y y^7 \partial_y + 1 +
    {64 (\omega R)^6 \over y^6} \right] \psi = 0 
  \ .}
 On can make the expansions
  \eqn{ExpandPhiPsi}{\eqalign{
   \phi &= \phi_0 + (\omega R)^3 \phi_1 + (\omega R)^6 \phi_2 + \ldots  \cr
   \psi &= \psi_0 + (2 \omega R)^6 \psi_1 + \ldots \ .  
  }}
 It is necessary to calculate only the terms shown explicitly in
(\ref{ExpandPhiPsi}):
  \eqn{CalculatedTerms}{\eqalign{
   \phi_0 &= {J_{3/2}(\rho) \over \rho^{3/2}} =
    \tf{1}{3} \sqrt{2 \over \pi} \left( 1 - {\rho^2 \over 10} + \ldots
     \right)  \cr
   \phi_0 &= -{\pi \over 2 \rho^{3/2}}
    \left[ \int^\rho {d\sigma \over \sqrt{\sigma}} \phi_0(\sigma)
     J_{3/2}(\sigma) N_{3/2}(\rho) - 
    \int^\rho {d\sigma \over \sqrt{\sigma}} \phi_0(\sigma)
     N_{3/2}(\sigma) J_{3/2}(\rho) \right] \cr
     &= {1 \over 3 \sqrt{2 \pi} \rho} 
       \left( 1 - \tf{1}{5} \rho^2 + \ldots \right)  \cr
   \phi_2 &= -{\pi \over 2 \rho^{3/2}}
    \left[ \int^\rho {d\sigma \over \sqrt{\sigma}} \phi_1(\sigma)
     J_{3/2}(\sigma) N_{3/2}(\rho) -
    \int^\rho {d\sigma \over \sqrt{\sigma}} \phi_1(\sigma)
     N_{3/2}(\sigma) J_{3/2}(\rho) \right] \cr
     &= {1 \over 6 \sqrt{2 \pi} \rho^2} 
       \left( 1 - \tf{1}{5} \rho^2 \log\rho + \ldots \right)  \cr
   \psi_0 &= {H^{(1)}_3(y) \over y^3} = 
     -{16 i \over \pi y^6} \left( 1 + {y^2 \over 8} + 
       {y^4 \over 64} + \ldots \right)  \cr
   \psi_1 &= -{\pi \over 2 y^3} \int^y {dx \over x^2}
     \psi_0(x) \left( J_3(x) N_3(y) - N_3(x) J_3(y) \right)  \cr
    &= {2 i \over 5 \pi y^{10}} \left( 1 + {y^2 \over 4} - 
         {y^4 \log y \over 64} + \ldots \right) \ . 
  }}
 In fact $\phi_0$ and $\phi_1$ can be given in closed form in terms of
trigonometric functions.

Now we can do the matching.  Define a radial variable $z$ which is
finite in the matching region:
  \eqn{ZDefsMore}{
   z = r/R\ , \qquad \rho = \omega R z\ , \qquad 
   y = {2 \omega R \over \sqrt{z}} \ .
  }
 We obtain
  \eqn{MatchEx}{\eqalign{
   \phi_{\bf I} &= y^6 \left( \psi_0 + (2 \omega R)^6 \psi_1 
      + \ldots \right)  \cr
     &= -{16 i \over \pi} \left[ 1 + 
    (\omega R)^2 \left( {1 \over 2z} - {z^2 \over 10} \right) + 
    (\omega R)^4 \left( {1 \over 4 z^2} - {z \over 10} \right) +
    \tf{1}{10} (\omega R)^6 \log \omega R + \ldots \right]  \cr
   {\phi_{\bf III} \over A} &= \phi_0 + (\omega R)^3 \phi_1 +
     (\omega R)^6 \phi_2 + \ldots \cr
    &= \tf{1}{3} \sqrt{2 \over \pi} \left[ 1 + 
    (\omega R)^2 \left( {1 \over 2z} - {z^2 \over 10} \right) + 
    (\omega R)^4 \left( {1 \over 4 z^2} - {z \over 10} \right) -
    \tf{1}{20} (\omega R)^6 \log \omega R + \ldots \right] \ .  
  }}
 A match is achieved to the order shown by setting
  \eqn{AMatch}{
   A = -24 i \sqrt{2 \over \pi} \left( 1 + \tf{3}{20}
    (\omega R)^6 \log \omega R \right) \ ,
  }
 which means that the absorption probability is
  \eqn{CrossCorrected}{
   P = {\pi\over 9} (\omega R)^9 
    \left( 1 - \tf{3}{10} (\omega R)^6 \log \omega R+ \ldots \right) \ .
  }
 The coefficient on the logarithm matches the numerical
analysis we have performed. 
To get the corresponding result for the M2-brane, we only need to
replace $\omega$ by $\omega/2$ in (\ref{CrossCorrected}).

\section{The world-volume approach}
\label{WorldVol}

The world-volume analysis, as usual, is much simpler than the
supergravity once one realizes what the relevant effects are.  We will
begin by considering a single D3-brane, where the action involves only
an abelian gauge field and its superpartners.  In the past
\cite{kleb,gukt}, it has been sufficient to generalize the single
brane calculation to multiple branes simply by multiplying
cross-sections by a factor of $N^2$ to account for the $N^2$ massless
gauge fields on a group of $N$ coincident branes.  This unexpected
simplification is due to supersymmetric non-renormalization theorems
that sometimes imply that the one-loop result is not renormalized
\cite{gkThree}. All the lowest order diagrams are contained in the
abelian theory, so we may think of the abelian calculation as the
leading term in the $g_{YM}^2 N$ expansion of the non-abelian
calculation. Even if the coefficient is a non-trivial function of
$g_{YM}^2 N$, it may approach a smooth infinite coupling limit,
similar to the factor of $3/4$ in the entropy calculation \cite{gkp}.

Here we are dealing with a
more complicated situation than that encountered in earlier
literature, because the operators
responsible for the effects we have seen in earlier sections are
dimension $8$, and to obtain their precise form one requires input
from the non-abelian DBI action literature \cite{Arkold,gw,Ark}.

For processes that are on-shell in the
world-volume theory, the only relevant coupling of the dilaton to the
brane is through the gauge field, as was first realized in
\cite{kleb}.  The action is 
  \eqn{DBIAct}{\eqalign{
   S_{\rm DBI} &= -T_3 \int d^4 x \, 
    \sqrt{-\det \left( g_{\mu\nu} + 
     {e^{-\phi/2} \over \sqrt{T_3}} F_{\mu\nu} \right)}  \cr
    &= \int d^4 x \, \sqrt{g} \left[T_3 - 
     \tf{1}{4} e^{-\phi} F^2 - \tf{1}{8} {e^{-2\phi} \over T_3}
      \left( F^4 - \tf{1}{4} (F^2)^2 \right) + \ldots \right]  
  }}
 where $g_{\mu\nu}$ is the induced Einstein metric on the brane, and
we have renormalized $F_{\mu\nu}$ to
give the photon propagator a pole of residue one.  We have introduced
a short-hand for Lorentz traces of $F_{\mu\nu}$:
  \eqn{TraceShort}{
   F^n = F_{\mu_1}{}^{\mu_2} F_{\mu_2}{}^{\mu_3} \ldots 
    F_{\mu_n}{}^{\mu_1}  \ .
  }
 To make expressions compact it is useful to define
  \eqn{TwoOs}{
   {\cal O}_4 = F^2\ , \qquad {\cal O}_8 = F^4 - \tf{1}{4} (F^2)^2 \ .
  }
 From (\ref{DBIAct}) one can obtain an expression for the trace of the
stress-tensor:
  \eqn{TraceT}{
   T^\mu_\mu = -\tf{1}{2} {e^{-2\phi} \over T_3} {\cal O}_8 \ .
  }
This operator is dimension $8$, suitable to match the scaling of the
fixed scalar cross-section with energy.

\begin{figure}[t]
\centerline{\psfig{figure=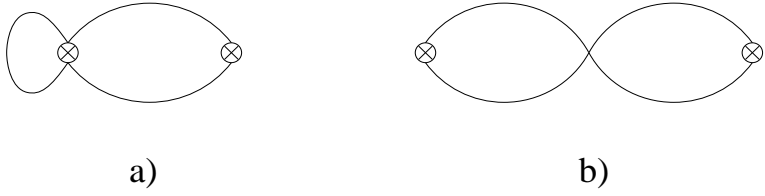,width=4in}}
\caption{The circled crosses ($\otimes$) denote an insertion of ${\cal O}_\phi$.\label{figA}}
\end{figure}

At linear order, the dilaton couples to the world-volume through
  \eqn{DilCoup}{
   S_{\rm int} = \int d^4 x \, \phi {\cal O}_\phi = 
    \int d^4 x \, \left( -\tf{1}{4} \right) \phi \left[ {\cal O}_4 +
    {1 \over T_3} {\cal O}_8 + \ldots \right] \ .
  }
 From the two-point function 
of the renormalizable part of ${\cal O}_\phi$ one
can read off the low-energy cross-section $\sigma = {\kappa^2 \omega^3
\over 32 \pi}$, which is correct even in normalization.  Our present
purpose is to consider the leading corrections to this two-point
function.  First of all, it is evident that in the expansion by
operator dimension, the higher one goes the more inverse powers of
$T_3$ show up in the correlators.  Extra powers of momenta (or just
energy in the case of normal incidence absorption cross-sections)
appear as required by dimensional analysis.  We are looking only for
the first correction, with a single inverse power of $T_3$ and four
powers of $\omega$, corresponding to $(\omega R)^4$ in supergravity.
To the order which we desire, then, 
  \eqn{PathApp}{\eqalign{
   \langle {\cal O}_\phi(x) {\cal O}_\phi(0) \rangle_{\rm DBI} &= 
    \int {\cal D} A_\mu \, 
     e^{-\int d^4 y \, \left[ {1 \over 4} {\cal O}_4 + 
      {1 \over 8 T_3} \, {\cal O}_8 \right]} {\cal O}_\phi(x)
      {\cal O}_\phi(0)  \cr
     &= \int {\cal D} A_\mu \, 
      e^{-\int d^4 y \, {1 \over 4} {\cal O}_4} \, {\cal O}_\phi(x)
      {\cal O}_\phi(0) \left( 1 - {1 \over 8 T_3} 
      \int d^4 z \, {\cal O}_8(z) \right)  \cr
     &= \left\langle {\cal O}_\phi(x) {\cal O}_\phi(0) 
      \left( 1 - {1 \over 8 T_3} \int d^4 z \, {\cal O}_8(z)
      \right) \right\rangle_{\rm CFT}  \cr
     &= \tf{1}{16} \left[ \langle {\cal O}_4(x) {\cal O}_4(0) 
         \rangle_{\rm CFT} - {1 \over 8 T_3} \int d^4 z \, 
        \langle {\cal O}_4(x) {\cal O}_8(z) {\cal O}_4(0)
         \rangle_{\rm CFT} \right]
   }}
 The subscripted CFT on the correlators in the last two lines of
(\ref{PathApp}) reminds us that these correlators are evaluated in the
free gauge theory, which of course is conformal.  From here on we will
drop this subscript: all correlators will be evaluated in the
conformal theory.  Note that we have gone to Euclidean signature in
order to simplify calculations.  In the last line of (\ref{PathApp})
we have suppressed terms like $\langle {\cal O}_4(x) {\cal O}_8(0)
\rangle$ coming from the dimension eight corrections to the vertex
operator (\ref{DilCoup}), which in principle seem to contribute at the
same order.  From a conformal field theory point of view, these terms
should vanish because only operators of the same dimension can have a
non-vanishing two point function.  From a diagrammatic point of view,
one can arrange for them to vanish by assuming (as we will below) that
${\cal O}_4$ and ${\cal O}_8$ are normal ordered.  Even if they were
not, their contributions would amount only to (divergent)
contributions to the absorption probability of order $(\omega R)^4$
without a log.  In other words, a full analysis of the pure $(\omega
R)^4$ correction might require us to consider diagrams like those in
figure~\ref{figA}a; but the leading log correction comes only from the figure
eight diagrams shown in figure~\ref{figA}b.

Now let us proceed to explicit evaluation of the free field
correlators. 
The relevant part of the action is
$${1 \over 4} (1 - \phi) F^2 + {1-2\phi \over 8 T_3} 
(F^4 - {1 \over 4}(F^2)^2)$$
In Euclidean space, and in Feynman gauge, the
propagator is
  \eqn{GaugeProp}{
   \langle A_\mu(x) A_\nu(0) \rangle = 
    {\delta_{\mu\nu} \over 4 \pi^2 x^2}}
from which one can infer the (gauge-invariant) propagator for the
field strength
 \eqn{FieldProp}{
   \langle F_{\mu\nu}(x) F_{\alpha\beta}(0) \rangle = 
    {1 \over 2 \pi^2 x^4} \left(
     \delta^{[\beta}_{[\alpha} \delta^{\nu]}_{\mu]} - 
     {4 \over x^2} \delta{^{[\beta}_{[\alpha}} x{_{\mu]}} x^{\nu]} 
    \right)
  }
Because we have rotated to Euclidean signature, there is no difference
between upper and lower indices. Using standard perturbative methods,
one finds
\begin{equation}
\Pi_0(x) = \langle \norm{{1 \over 4}F^2(x)} \norm{{1 \over 4}F^2(0)} \rangle = {3 \over \pi^4}{1 \over x^4}
\end{equation}
which gives the leading contribution, and
\begin{equation}
\Pi_1(x) = -\int d^4 z \, \langle \norm{{1 \over 4} F^2(x)} 
\norm{{1 \over 8 T_3}(F^4 - { 1 \over 4}(F^2)^2)(z)} 
\norm{{1 \over 4} F^2(0)} \rangle  = 
- \int d^4 z\, {9 \over \pi^8 T_3} {1 \over z^8 (x-z)^8} \label{cor484}
\end{equation}
which gives the correction arising from the ``figure 8'' diagram of
figure~\ref{figA}b. This form follows from operator product expansions
$$ {\cal O}_4(z) {\cal O}_4(0) = {1 \over z^8}{\bf 1} +  {\cal O}_8(0) + \ldots, \qquad 
{\cal O}_8(z) {\cal O}_4(0) = {1 \over z^8} {\cal O}_4(0) + \ldots
$$
In momentum space, they become
\begin{eqnarray*}
\Pi_0(p) &=& \int d^4x\, \Pi_0(x) e^{ipx} = 
{p^4  \over 64 \pi^2} \log(p^2/\Lambda^2) \\
\Pi_1(p) &=& \int d^4x\, \Pi_1(x) e^{ipx} = - {p^8 \over 4096 \pi^4 T_3}
\left( \log(p^2/\Lambda^2) \right)^2
\end{eqnarray*}
As was explained in \cite{gkThree}, absorption cross section can be
inferred from the discontinuity across the positive real axis in the
$s$-plane ($s = -p^2$) of $\Pi_0$ and $\Pi_1$.
$$\sigma = {2 \kappa^2 \over 2 i \omega} \left.\disc 
(\Pi_0(s) + \Pi_1(s))\rule{0ex}{2ex} 
\right|_{{p^0 = \omega  \atop \vec{p} =0}} =  
{\kappa^2 \omega^3  \over 32 \pi}\left( 1  
-  {1 \over 8} (\omega R)^4 \log(\omega / \Lambda)\right)
$$
where we used the  relation between 
radius of the throat $R$ and the 3-brane tension $T_3$,
$$R^4 ={N \over 2 T_3 \pi^2} = 4 \pi g N \alpha'^2.$$
This is precisely the form of the dependence on $\omega$ we found in
the previous section using supergravity methods. The coefficient of
the $\log(\omega)$ term disagrees from the supergravity calculation by
a factor of $4/3$.  This is not necessarily a contradiction, however.
One must keep in mind that we trust the supergravity calculation in
the limit where $g^2_{YM} N$ is large.  From the world volume point of
view, one must take proper account of the non-abelian nature of the
theory.  This might lead to a non-trivial dependence on $g^2_{YM}N$
which approaches a smooth limit as $g^2_{YM}N$ is sent to infinity,
just as in the case of the entropy calculation \cite{gkp}.

In the full non-abelian theory the calculation of the correction to
the absorption cross-section once again reduces to the correlator
$\langle {\cal O}_4(x){\cal O}_8(z){\cal O}_4(0)\rangle$, except now
we use the non-abelian operators given in (\ref{dimfour}) and
(\ref{irrop}). Due to the constraints of conformal invariance, the
3-point function is, up to an overall function of $g^2_{YM} N$, the
same as at the leading order in $g^2_{YM} N$, 
\eqn{NonAb} {\langle
{\cal O}_4(x){\cal O}_8(z){\cal O}_4(0)\rangle \sim {1\over z^8
(x-z)^8} \ .}  
According to the discussion above, this guarantees that
the absorption cross-section has the form
$${\kappa^2 \omega^3  \over 32 \pi}\left( 1  
-  f(g^2_{YM} N) \omega^4 N \kappa\log(\omega / \Lambda)\right)
\ .$$
However, conformal invariance alone does not fix the function
$f(g^2_{YM} N) $. Our perturbative calculation
tells us that $f(0)=1/8$.
On the other hand, since the supergravity calculation is reliable in
the limit of large $g^2_{YM} N$, we can use (\ref{CrossSRatio}) to
deduce that $f=1/6$ in this limit.  This in turn fixes the
normalization of the 3-point function in (\ref{NonAb}). Of course, a
detailed understanding of the strong coupling limit
of the world-brane theory will be required to check it.

A comment is in order regarding the ultraviolet cut-off scale
$\Lambda$.  In order to match the supergravity result
(\ref{CrossSRatio}), one should set $\Lambda = 1/R$.  On the other
hand, these world volume theories are an effective description of open
strings ending on a D-brane, and the natural ultraviolet cut-off is
$\Lambda = 1 / \sqrt{\alpha'}$. However, for arbitrary choice of the
cut-off scale $\Lambda$, supergravity and world volume calculations
differ only by a quantity of order
$$ (\omega R)^4 \log(\omega R) - (\omega R)^4 \log(\omega /\Lambda) =
(\omega R)^4 \log(\Lambda R) $$
which can be thought of as being part of the $(\omega R)^4$ term
(without the logarithm) which is subleading. Proper understanding of
the ultraviolet cutoff therefore requires extending our analysis to
this order.  It would be interesting to carry out this
analysis, perhaps by looking at the string theory amplitudes. We leave
this point for future investigations.

A similar story appears to hold for the M2 and M5 branes.  For the
M5-brane, minimal and fixed scalars couple to operators of dimensions
6 and 12, respectively. Assuming that the the operator product
expansion are of the form
$$ {\cal O}_6(z) {\cal O}_6(0) = 
{1 \over z^{12}}{\bf 1} +  {\cal O}_{12}(0) + \ldots, \qquad 
{\cal O}_{12}(z) {\cal O}_6(0) = {1 \over z^{12}} {\cal O}_6(0) + \ldots
$$
conformal invariance will fix their correlation functions to take the form
$$
\Pi_0(x) = {1 \over x^{12}}, \qquad
\Pi_1(x) = \int d^6 z\  {1 \over z^{12} (x-z)^{12}}
$$
On the $s$-space, they become
$$
\Pi_0(s) = s^3 \log(-s), \qquad
\Pi_1(s) = s^6 \log(-s)^2
$$
whose discontinuity on the real axis is
$$\disc (\Pi_0(s) + \Pi_1(s)) = i\, s^3 (1 + s^3 \log(s))$$
which gives the desired $(\omega R)^6 \log(\omega / \Lambda)$
dependence of the sub-leading term.

The pattern that appears to be emerging is that when minimal scalars
couple to the operator on the brane world volume of dimension $d$,
there is a leading logarithmic correction at order $(\omega R)^d
\log(\omega/\Lambda)$. On the M2-brane, minimal scalars couple to
operators of dimension 3. This appears to suggest that there is a
logarithmic correction of order $(\omega R)^3 \log(\omega / \Lambda)$
which would contradict our earlier result from supergravity calculation,
where the leading correction to the minimal scalar absorption was
found to be of order $(\omega R)^6 \log(\omega / \Lambda)$.  Quite
happily, we find that the conformal field theory on M2 knows about the
absence of $(\omega R)^3 \log(\omega/\Lambda)$ correction upon closer
examination. Let us elaborate on this point.

\begin{figure}[t]
\centerline{\psfig{figure=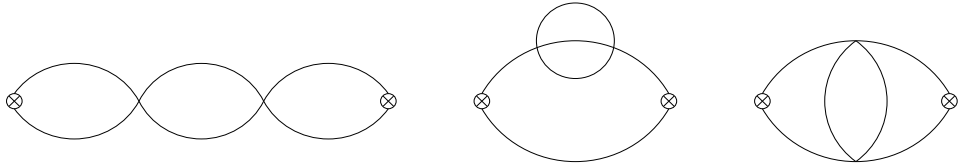,width=5in}}
\caption{The three loop contributions to the two point function.
 \label{figB}}
\end{figure}

The dimension of operators on M2 world volume coupling to the minimal
and fixed scalars are 3 and 6, respectively. Assuming that the the
operator product expansion are again of the form
$$ {\cal O}_3(z) {\cal O}_3(0) = 
{1 \over z^{6}}{\bf 1} +  {\cal O}_{6}(0) + \ldots, \qquad 
{\cal O}_{6}(z) {\cal O}_3(0) = {1 \over z^6} {\cal O}_3(0) + \ldots
$$
conformal invariance will fix their correlation functions to take the form
$$
\Pi_0(x) = {1 \over x^{6}}, \qquad
\Pi_1(x) = \int d^3 z\  {1 \over z^{6} (x-z)^{6}}
$$
This time, we find that on  $s$-space, they become
$$
\Pi_0(s) = (-s)^{3/2}, \qquad
\Pi_1(s) = s^3.
$$
The leading non-analytic behavior is a square root instead of the logarithm. 
In particular, $\Pi_1(s)$ has no non-analyticity what so ever, and we find
\eqn{TwoDisc}{\disc (\Pi_0(s) + \Pi_1(s)) = i\, 
s^{3/2} }
 which gives the desired leading term, and the desired absence of
$(\omega R)^3 \log(\omega / \Lambda)$ correction.  Power counting in
$\omega$ and $R$ indicates that the desired $(\omega R)^6 \log(\omega
/ \Lambda)$ correction should arise from the next correction to the
two point function:
  \eqn{FourPointCor}{
   \Pi_2(x) = \int d^3 z_1\, d^3 z_2 \langle {\cal O}_3(x) {\cal O}_6
   (z_1) {\cal O}_6 (z_2) {\cal O}_3 (0) \rangle 
  }
 To see what singularities arise from (\ref{FourPointCor}) it is
useful to replace the interacting CFT of the M2 world-volume with a
free CFT: for example a free conformally coupled scalar $\phi$.  Then,
roughly speaking, ${\cal O}_3$ could be realized as $(\partial\phi)^2$
and ${\cal O}_6$ as $(\partial\phi)^4$.  The three diagrams arising
from the Wick contractions of (\ref{FourPointCor}) are shown in
figure~\ref{figB}.  The first two only contribute $s^{9/2}$
corrections to (\ref{TwoDisc}); the last however contains overlapping
divergences and can contribute a $s^{9/2} \log(s)$ term to
(\ref{TwoDisc}) through the Fourier transform of integrals like
  \eqn{CateyeInt}{
   \int d^3 z_1 d^3 z_2 \, {1 \over (x-z_1)^3} {1 \over (x-z_2)^3}
    {1 \over z_1^3} {1 \over z_2^3} {1 \over (z_1-z_2)^6} \sim
    {\log \Lambda x \over x^{12}} \ .
  }
 A coupling of the minimal scalar to ${\cal O}_6$ does not seem to
give additional logarithmic contributions.

\section{Conclusions}
\label{Conclude}

The proposal \cite{jthroat} that the world-volume theory of a 
large number of coincident branes should 
be identified with string theory in the background of the
near-horizon geometry is an elaboration of earlier work 
\cite{kleb,gukt,gkThree} where the entire brane geometry (which
goes beyond the throat approximation) was used. 
In this paper we studied some of the effects that go beyond 
the throat approximation and argued that they are just as universal as the
strict conformal limit.

We have presented two closely related calculations. First, we identified
a non-minimally coupled ``fixed'' scalar
and showed that a coupling of this scalar to the trace of the
stress-energy tensor of the full DBI action reproduces the
supergravity cross-section up to a normalization.  We regard the
normalization question in this case as something of a fine point: as
in comparisons of matrix models to Liouville theory, the
 things to compare between the world-volume theory and
supergravity are ratios of two and three point functions which are
independent of normalizations.

The calculation of logarithmic corrections to the energy dependence of
the minimal scalar cross-section strikes more deeply at the physics of
D-branes: whereas the world-volume calculation for fixed scalar
required the DBI action only in order to find the correct vertex
operator for the fixed scalar, the logarithmic corrections to minimal
scalar cross-sections arise from considering DBI corrections to the
conformal theory in the path integral.  

One striking feature of our results is that the semi-classical supergravity
is being compared with the $\alpha'$ corrections to the DBI
action, always thought to be a purely string theoretic effect.
Thus, we are finding out that the supergravity `knows' string
theory.

Our calculations suggest an extension of the throat-brane
correspondence conjecture, which was implicit in the earlier work of
\cite{kleb,gukt,gkThree}: the world-volume theory of $N$ coincident
D3-branes interacting with closed strings in the bulk is equivalent to
type IIB string theory in the background of the self-dual black
threebrane carrying $N$ units of RR charge.  One can make a more
concrete claim at low energy: expanding departures in the world-volume
theory from conformal invariance in a series of irrelevant
perturbations to the Yang-Mills action corresponds to considering
supergravity in the throat region, but with the usual $AdS$ throat
metric replaced by the true D3-brane metric.  In this paper we
compared the leading correction to the absorption cross-section in the
double-scaling limit (\ref{dsl}) proposed in \cite{kleb}.\footnote{It
would be very interesting to see if the higher order corrections can
be compared explicitly.  In fact, we conjecture that the
semi-classical absorption cross-section calculated as a function of
$(\omega R)^4$ from the wave equation (\ref{minmassless}) defines a
``universal function'' which should be reproduced by the non-abelian
DBI theory in the double-scaling limit (\ref{dsl}).} Our calculation
of the corrections due to the first irrelevant perturbation, together
with the work on BIons in \cite{jCal,bion,lpt,rey}, perhaps constitute
only the beginning of a fuller understanding of the relation between
the non-abelian DBI action and string theory in the brane background.

\section*{Acknowledgement}

We would like to thank A. Tseytlin and the participants of the
String Dualities program at ITP, Santa Barbara, for 
discussions. S.S.G. and I.R.K. are grateful to ITP for
hospitality during the completion of this paper.
This research was supported in part by the National Science
Foundation under Grant No. PHY94-07194, by the Department
of Energy under Grant No.
DE-FG02-91ER40671,
and by the James S.~McDonnell Foundation under Grant No. 91-48.
S.S.G. also thanks the Hertz Foundation for its support.

\bibliography{imp}
\bibliographystyle{ssg}

\end{document}